\documentclass[conference,10pt]{IEEEtran}

\usepackage{amssymb,amsmath,amsfonts,bm,epsfig,graphicx,theorem,latexsym}
\usepackage{rotating,setspace,latexsym,epsf,color,subfigure}
\usepackage{cite,authblk,float,epsf,bbm}

\def \n2{{N_0 \over 2}}

\def \h5{\hspace{0.5in}}

\begin{document}
\IEEEoverridecommandlockouts
\pagestyle{empty}

\title{Maintaining Information Freshness in Power-Efficient Status Update Systems\vspace{-0.1in}}

\author{Parisa Rafiee \quad Peng Zou \quad Omur Ozel \quad Suresh Subramaniam \\
\normalsize Department of Electrical and Computer Engineering \\
\normalsize George Washington University, Washington, DC 20052 USA \\
\normalsize {\it \{rafiee, pzou94, ozel, suresh\}@gwu.edu}\vspace{-0.15in}}

\maketitle 

\begin{abstract}
This paper is motivated by emerging edge computing systems which consist of sensor nodes that acquire and process information and then transmit status updates to an edge receiver for possible further processing. As power is a scarce resource at the sensor nodes, the system is modeled as a tandem computation-transmission queue with power-efficient computing. Jobs arrive at the computation server with rate $\lambda$ as a Poisson process with no available data buffer. The computation server can be in one of three states: (i) OFF: the server is turned off and no jobs are observed or processed, (ii) ON-Idle: the server is turned on but there is no job in the server, (iii) ON-Busy: the server is turned on and a job is processed in the server. These states cost zero, one and $p_c > 1$ units of power, respectively. Under a long-term power constraint, the computation server switches from one state to another in sequence: first a deterministic $T_{o}$ time units in OFF state, then waiting for a job arrival in ON-Idle state and then in ON-Busy state for an independent identically distributed compute time duration. The transmission server has a single unit data buffer to save incoming packets and applies last come first serve with discarding as well as a packet deadline to discard a sitting packet for maintaining information freshness, which is measured by the Age of Information (AoI). Additionally, there is a monotonic functional relation between the mean time spent in ON-Busy state and the mean transmission time. We obtain closed-form expressions for average AoI and average peak AoI. Our numerical results illustrate various regimes of operation for best AoI performances optimized over packet deadlines with relation to power efficiency. 
\end{abstract}

\section{Introduction}

Emerging edge computing systems consist of a large number of sensor nodes which continually acquire and process data before transmitting update packets to an edge receiver for further processing. These sensor nodes are typically power-constrained. In this paper, we use the Age of Information (AoI) metric to analyze a tandem computation-transmission queue and the tradeoff between the two operations for time-critical information updating. AoI is now a widely studied metric as a measure of staleness of status updates at monitoring receivers of a system and our goal is to use it for the same with additional power efficiency considerations in the computation. Since the publication of \cite{kaul2012status, kaul2012real} that pioneered AoI analysis for various queuing models, the literature on AoI and its applications have expanded considerably. \cite{costa2016age} investigates the role of packet management to improve the average AoI at the monitoring node. \cite{inoue2018general} provides a general treatment of stationary probability analysis of AoI in various preemptive and non-preemptive queuing disciplines. Reference \cite{kam2018age} considers introducing packet deadlines to discard the packets in a single server system for improving average AoI. In \cite{infocom_arxiv,infocom_w} \textit{waiting} is used as a mechanism to regulate the traffic while in \cite{pimrc19} tandem computation-transmission operations and queue management are combined. We also refer the reader to \cite{najm2017status,sun2017update, bedewy2017age, talak2017minimizing, yates2018age, maatouk2018age,Gong2019ReducingAF, bacinoglu2015age, yates2015lazy, wu2017optimal_ieee, arafa2017age,farazi2018average, feng2018optimal, bacinoglu2018achieving, leng2019age,rafiee19} for other work closely related to this paper. 

In applications where the main power consumption is due to computation, a natural power-efficient scheme is to intermittently take the system to a stand-by mode with minimal power consumption to save power. There is an unavoidable risk of missing time-critical information during stand-by, and therefore a need to understand the tradeoff between power-saving and timeliness. This paper aims to shed light on this tradeoff in a tandem computation-transmission queue. We build on previous papers \cite{infocom_arxiv,infocom_w,pimrc19} and perform average AoI and average peak AoI analysis with two new aspects that were not considered before: (a) The computation server goes to OFF (stand-by) state as a means to save power, and (b) The transmission server applies a packet deadline for the sitting packet while using a last come first serve (LCFS) with discarding policy. As in \cite{pimrc19}, our motivation comes from applications in which computation at the sensor node could be prolonged in order to reduce the amount of data to be transmitted to the edge receiver. For example, the computation unit could represent an image processing device that has to inform a remote receiver about the image it captured in a timely manner. The more processing the device performs on the image before transmission, the less amount of work remains to be done elsewhere. This aspect of the problem is modeled as a monotone decreasing relation between the time spent in ON-Busy state and the time spent for transmission.

\begin{figure}[!t]
\centering{
\hspace{-0.0cm} 
\includegraphics[totalheight=0.10\textheight]{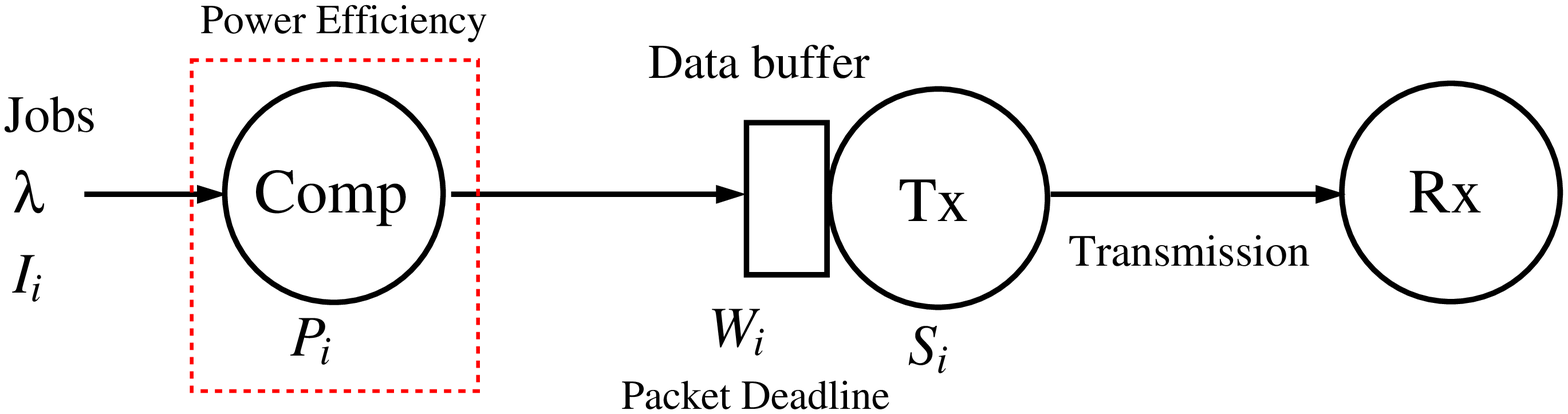}}\vspace{-0.1in}
\caption{\sl System model for our status update system with tandem computation-transmission queues having power efficient computing and communication queue applying a packet deadline.}\vspace{-0.25in}
\label{fig:1} 
\end{figure}

In our model, a job arrives into the tandem computation-transmission queue shown in Fig. \ref{fig:1} as a Poisson process. If the job finds the server in OFF or ON-Busy states, it is discarded; otherwise, it is taken to the server for processing and then it is sent to the transmission queue. Processing time is independent identically distributed (i.i.d.) and it has a general distribution. \textit{Neither the transmission nor the computation services are preemptive.} The transmission service time is exponentially distributed. A single unit data buffer is available for the transmission server where a last come first serve with discarding policy is applied and a packet that sits in the buffer longer than a threshold level is discarded to maintain information freshness. We are particularly motivated by \cite{kam2018age,ozel2020} to introduce packet deadlines for sitting packets in the data buffer. We determine closed form expressions for average AoI and average peak AoI. We provide performance comparisons to show various operating regimes.

\section{System Model}
\label{sec:Model}

We consider a computation server followed by a transmission queue in tandem as shown in Fig. \ref{fig:1}. Here, computation represents initial operations such as image capture, sensing, compression needed to generate a status update packet to be sent to a remote receiver. We assume that jobs (which we think to be initiated by events of interest) arrive into the computation server as a Poisson process of rate $\lambda$. The aging of the jobs starts as soon as the computation starts. Once the computation is completed, a status update packet is fed into the communication queue where there is a single data buffer for the latest arriving packet while any sitting packet is discarded.

\subsection{Power Efficient Computing}

The computation server has three states: OFF, ON-Idle and ON-Busy. The server is designed to change from one state to the other in a sequential way as follows: First, it remains in OFF state for $T_{o}$ unit time. Then, it switches to ON-Idle, waiting in idle state for a job to arrive. Finally, a job arrives and the system is taken to ON-Busy state. The server spends $0$, $1$ and $p_c>1$ units of power in these states, respectively.

Computation time $P_i$ for job $i$ to be processed in ON-Busy state is independent identically distributed over $i$ with a general distribution $f_P(p)$, $p \geq 0$ with mean $\mathbb{E}[P]$. We denote the moment generating function (MGF) of the computation time at $-\gamma$ with $M_{P}(\gamma)$:
\begin{align}
M_{P}(\gamma) = \mathbb{E}[e^{-\gamma P}].
\end{align}

There is a long-term average constraint $C_{avg}$ on the power cost incurred in the computation server. Due to the renewal structure, a renewal cycle has average length $T_{o} + \frac{1}{\lambda} + \mathbb{E}[P]$ and the average power consumption constraint is:
\begin{align}\label{const}
\frac{\frac{1}{\lambda} + p_c \mathbb{E}[P]}{T_{o} + \frac{1}{\lambda} + \mathbb{E}[P]} \leq C_{avg}.
\end{align} 
This constraint on average power consumption can equivalently be expressed as a linear constraint on $\mathbb{E}[P]$ as follows: \\ $(p_c - C_{avg}) \mathbb{E}[P] \leq C_{avg}T_o +\frac{1}{\lambda}(C_{avg} - 1)$.

\subsection{Packet Deadlines}

The transmission server applies a packet deadline $\tau$: If the packet sits in the data buffer for longer than $\tau$, it is discarded.\footnote{An alternative way of implementing packet deadline would be to include the initial time $P_i$ spent in the computation server. Since we view the mean value of the computation time as a parameter of the whole system to be optimized, we do not follow this approach in this work.} This discarding is done in the same spirit as \cite{kam2018age,ozel2020} where usefulness of keeping young packets in the data buffer are shown. Transmission is performed one at a time and its duration is a random variable with an exponential distribution with rate $\mu$. Once transmission is completed, the monitoring receiver (Rx) has the most recent update. Both computation and transmission services are non-preemptive. We finally assume that the threshold applied in the communication queue $\tau$ is smaller than the OFF time $T_{o}$ to ensure that the sent packet to the receiver is reasonably \textit{young:}  \[ \tau \leq T_o \]

\subsection{Dependence of Mean Transmission and Processing Times}

Typical to edge computing applications is an inverse relationship between computation time spent to generate the packet and the time to transmit the resulting packet to a remote computer (where the time-sensitive application is run). For example, in the case of a new captured image to be processed, some portion of the processing is done in the device that captures it, which then delegates the remaining work to the remote computer. We address this characteristic in our model in that the mean compute time in ON-Busy state $\mathbb{E}[P]$ and mean transmission time $\mathbb{E}[S]=\frac{1}{\mu}$ are dependent through a monotone decreasing function $g$ as:
\begin{align}
\frac{1}{\mu} = g(\mathbb{E}[P]).
\end{align}
This monotonic dependence along with the power constraint in (\ref{const}) causes a conflicting situation where on one side computing longer is expected to decrease the load on the transmission queue while incurring larger power cost.

\subsection{Equivalent Queue Model}

We use the equivalent queue model approach in \cite{infocom_w, infocom_arxiv,pimrc19} to analyze the system. This model produces identical AoI expression as in the actual system and simplifies analysis. In this model, all arriving packets to the transmission queue are stored in the queue and served; and, the data buffer capacity is unlimited. Different from the original model, multiple packets are served at the same time in the second queue. An arriving packet may find the transmission queue in Idle (Id) or Busy (B) states. If a packet finds the second queue in (Id), then that packet enters the server at that instant; if the queue is in (B) state, the packet enters the server after the end of the current service period. All the packets arriving in state (B) enter the server altogether as a batch at the same instant when the ongoing transmission ends. No modification is required in the computation queue.

We let $t_i$ and $t_i'$ denote the time stamps of the event that job $i$ enters the computation queue and the event of completion of delivery to the receiver. We count only those that enter the queue and assume no packet is generated while the computation server is in ON-busy or OFF states. Each index $i$ represents a job processed in the computation and an update packet entering the transmission queue. The age is counted starting from the instant a packet enters the computation server. 

Denoting the inter-arrival time between two successive jobs $i-1$ and $i$ entering the computation queue as $X_i$, we observe that $X_i = P_{i-1} + T_{o} + I_i$ where $I_i$ is independent exponentially distributed idle time with rate $\lambda$. Additionally, $T_i = P_i + W_i + S_i$ is the system time for packet $i$ starting from its entry to the computation server until its delivery to the receiver. In here, $W_i \geq 0$ is the length of time packet $i$ sits in the data buffer of the transmission queue before being taken to the server. Recall that the server applies a deadline and if $W_i$ is larger than $\tau$, then it is discarded in the original model. In the equivalent model, if $W_i$ is larger than $\tau$, then it is served together with the next arriving packet. $S_i$ is the transmission service time for packet $i$. Age of information (AoI) is the difference of current time and the time stamp of the packet at the receiver: \begin{align} \Delta(t)=t-u(t) \end{align} where $u(t)$ is the time stamp of the latest packet at the receiver at time $t$. We express $u(t)=t_{i^*}$ where $i^*=\max\{i: \ t_i' \leq t\}$. 

At this point, we refer the reader to our earlier work \cite{pimrc19} as the definitions of $X_i$ and $T_i$ in the equivalent queuing model closely follow those in \cite{pimrc19}. Since the equivalent queue is in first come first serve form, we have the average AoI:
\begin{equation}\label{aaoi}
\mathbb{E}[\Delta]=\widetilde{\lambda}\left(\mathbb{E}[XT]+\frac{\mathbb{E}[X^{2}]}{2}\right),
\end{equation}
where $\widetilde{\lambda} = \frac{\lambda}{ \lambda \mathbb{E}[P] + \lambda T_{o} + 1}$ is the effective arrival rate for the system. Additionally, we have
\begin{align}\label{ap1}
\mathbb{E}[X^2]=\mathbb{E}[P^2] + 2\mathbb{E}[P](\frac{1}{\lambda}+T_{o}) + \frac{2}{\lambda^2} + T_{o}^2 + 2\frac{T_{o}}{\lambda}.
\end{align}
It then remains to calculate $\mathbb{E}[XT]$ to obtain average AoI. We will also get average peak AoI $PAoI_{i^*}$ as in \cite{pimrc19} by finding the maximum $X_j+T_j$ among all packets $j$ served during a service period and $i^*$ is the smallest index among all of them. 

\section{Average AoI and Average Peak AoI}
\label{sec:eval}

When packet $i$ leaves the computation server, the transmission queue can be in (Id) or (B) states: $K_i \in \{ (Id), (B)\}$. 
We note that $K_i$ is a two-state Markov chain. Conditioned on $K_{i-1}=(Id)$, $K_{i}=(Id)$ iff $T_{o} + I_i + P_i > S_{i-1}$. On the other hand, conditioned on $K_{i-1}=(B)$, packet $i-1$ observes a non-zero residual service time $R_{i-1}$ and $K_{i}=(B)$ under two conditions: (i) If $R_{i-1}>\tau$ and $T_{o}+ I_i + P_i < R_{i-1}$, then the residual transmission time packet $i-1$ observes at the time of arrival still continues (while packet $i-1$ has already been discarded due to long wait in the data buffer). Recalling that $\tau \leq T_{o}$, we can simply state this condition as $T_{o}+ I_i + P_i < R_{i-1}$; (ii) If $T_{o} + I_i + P_i < R_{i-1} + S_{i-1}$ and $R_{i-1}<\tau$, then packet $i-1$ is in service when packet $i$ arrives and hence packet $i$ finds the queue in (B) state. Note that both $R_i$ and $S_i$ are exponentially distributed with rate $\mu$ and they are independent variables. This way, $K_i$ is a two-state Markov chain with transition probabilities:
\begin{align}
&\hspace{-0.13in}\mbox{Pr}[K_{i}=(B)|K_{i-1}=(Id)]=\mbox{Pr}[T_{o} + I_i + P_i < S_{i-1}],\\ \nonumber
&\hspace{-0.13in}\mbox{Pr}[K_{i}=(B)|K_{i-1}=(B)]= \mbox{Pr}[T_{o} + I_i + P_i < R_{i-1}] \\ &+\mbox{Pr}[T_{o} + I_i + P_i < R_{i-1} + S_{i-1},\ R_{i-1}<\tau].
\end{align} 
To calculate the first of these transition probabilities, we note
\begin{align*}
\hspace{-0.1in}\mbox{Pr}[T_{o}+I_i + P_i < S_{i-1}] = \mathbb{E}[e^{-\mu (T_{o}+I_i + P_i)}] =\frac{\lambda e^{-\mu T_{o}}}{\lambda + \mu} M^{P}(\mu).
\end{align*}
Next, we attempt to calculate the second probability. We have, for the first term, $\mbox{Pr}[T_{o}+I_i + P_i < S_{i-1}]=\mbox{Pr}[T_{o}+I_i + P_i < R_{i-1}]=\frac{\lambda e^{-\mu T_{o}}}{\lambda + \mu} M^{P}(\mu)$. We then have for the second term:
\begin{align*}\nonumber
\mbox{Pr}&[T_{o} + I_i + P_i < R_{i-1} + S_{i-1},\ R_{i-1}<\tau] \\ &= \mathbb{E}_{I,P}[\int_0^\tau \int^{\infty}_{T_{o}+I+P - R} \mu e^{-\mu s}ds \mu e^{-\mu r}dr] \\ &= \mathbb{E}_{I,P}[ \tau \mu e^{- \mu (T_{o} + I + P)}] = \tau \mu \frac{\lambda e^{-\mu T_{o}}}{\lambda + \mu} M^{P}(\mu), 
\end{align*}
where we use $\tau \leq T_{o}$. Then, the stationary probabilities are
\begin{align}
p_B=\frac{\lambda e^{-\mu T_{o}} M^{P}(\mu)}{\lambda + \mu -\lambda e^{-\mu T_{o}} \tau\mu M^{P}(\mu)}, \label{pb}
\end{align}
and $p_I = 1- p_B$, where we use $p_B=\mbox{Pr}[K_i = (B)]$ to denote the stationary probability of being in (B).  

\subsection{Average AoI}

We now evaluate $\mathbb{E}[XT]$ by following \cite{pimrc19} to obtain conditional expectations under $K_{i-1}=(Id)$ and $K_{i-1}=(B)$.

\subsubsection{$\mathbb{E}[X_iT_i \ | \ K_{i-1}=(Id)]$}

In this case, packet $i-1$ finds the second queue in (Id) state. $X_i = P_{i-1} + T_{o} + I_{i}$ and if $T_{o} + I_i + P_i > S_{i-1}$, then packet $i$ is taken to service right away and $T_i = P_i + S_i$. On the other hand, if $T_{o} + I_i + P_i < S_{i-1}$, then $T_i = P_i + W_i + S_i$ and we next characterize wait time $W_i$. Note that there will be a non-zero residual time packet $i$ observes, which we denote by $R_i$. If $R_i$ is larger than $\tau$, then the packet is discarded and arrival of the next packet is awaited. When the next packet arrives, if it observes a residual time larger than $\tau$, that packet is also discarded. This process continues until a packet is ensured to remain \textit{young} at the end of transmission. In the equivalent model, we assume that packet $i$ and all potential future packets that find the queue in (B) state remain in the buffer and they are served together once such a \textit{young} packet is found. So, if $R_i \leq \tau$, then $T_i = P_i + R_i + S_i$. If $R_i>\tau$ and $T_{o} + I_{k_1} + P_{k_1} > R_i$ (where $k_1$ represents the very next trial), then we have $T_i = P_i + T_{o} + I_{k_1} + P_{k_1} + S_i$. If $T_{o} + I_{k_1} + P_{k_1} < R_i$ and $R_i - (T_{o} + I_{k_1} + P_{k_1}) \leq \tau$, then $T_i = P_i + T_{o} + I_{k_1} + P_{k_1} + \tilde{R} + S_i$ where $\tilde{R} \leq \tau$ is the residual time under assumed condition (which is also exponentially distributed with rate $\mu$). We conclude that the wait time of packet $i$ is in the following form
\begin{equation}
\label{frm}
\hspace{-0.12in}W_i=\left\{\begin{array}{ll}
\sum_{j=1}^{\tilde{n}} T_{o} + I_{k_j} + P_{k_j}, & \mbox{for condition (a)}\\ M + \sum_{j=1}^{\tilde{n}} T_{o} + I_{k_j} + P_{k_j}, & \mbox{for condition (b)},
\end{array}
\right.
\end{equation}

\noindent where we refer to the following conditions: \vspace{0.1in} \\
\vspace{0.1in} (a) $T_{o} + I_{k_{\tilde{n}}} + P_{k_{\tilde{n}}}> R_i - \sum_{j=1}^{\tilde{n}-1}(T_{o} + I_{k_j} + P_{k_j}) >\tau$ \\ 
\vspace{0.1in} (b) $0 \leq R_i - \sum_{j=1}^{\tilde{n}}(T_{o} + I_{k_j} + P_{k_j}) \leq \tau$. \\
Additionally, $\tilde{n}$ is the corresponding stopping time and $M$ in condition (b) is residual time $0 \leq M \leq \tau$ with exponential distribution of rate $\mu$ restricted to $[0,\tau]$. Summations involving $0$ in the upper limit are assumed $0$. As a special case, only condition (b) is checked with $\tilde{n}=0$ and if satisfied then the process stops and $M=R_i$; otherwise, for $\tilde{n}>0$ both conditions are checked. Here, $\tilde{n}$ has the following distribution: $\mbox{Pr}(\tilde{n}=0)=\mbox{Pr}(R_i\leq \tau) = 1-e^{-\mu \tau}$ and conditioned on $\tilde{n}>0$, it has a geometric distribution with success probability $\mbox{Pr}(R \leq T_{o} + I + P)=1-\frac{\lambda e^{-\mu T_{o}}}{\lambda + \mu} M^{P}(\mu)$ where $R$, $I$, and $P$ are independent random variables with $R$, $I$ having exponential distribution of rate $\mu$, $\lambda$ (resp.) and $P$ having density $f_{P}(p)$. Finally, note that $\mbox{Pr}(R \leq T_{o} - \tau + I + P)=1-\frac{\lambda e^{-\mu (T_{o}-\tau)}}{\lambda + \mu} M^{P}(\mu)$ for $\tau \leq T_o$ and hence, we get the probability of observing condition (a) when stopped is
\[\mbox{Pr}[ (a) ] = \frac{\lambda + \mu - \lambda e^{-\mu (T_{o}-\tau)} M^{P}(\mu)}{\lambda + \mu - \lambda e^{-\mu T_{o}} M^{P}(\mu)},\] 
and $\mbox{Pr}[ (b)] = 1 - \mbox{Pr}[ (a)]$. We then get the following:
\begin{align*}
\mathbb{E}[W_i]&=(1-e^{-\mu \tau})\mathbb{E}[R_i | R_i \leq \tau]  \\ &\ + e^{-\mu\tau}(\mathbb{E}[\tilde{n}|\tilde{n}>0]\mathbb{E}[T_{o} + I_{k} + P_{k}] +\mathbb{E}[M|\tilde{n}>0]) \\ &= \Big((1-e^{-\mu \tau})+ e^{-\mu\tau}\mbox{Pr}[ (b) ] \Big)(\frac{1}{\mu} - \frac{\tau e^{-\mu\tau}}{1-e^{-\mu \tau}}) \\ &\ + e^{-\mu\tau}\frac{\lambda + \mu}{\lambda+\mu-\lambda e^{-\mu T_{o}}M^{P}(\mu)}(T_{o} + \frac{1}{\lambda} + \mathbb{E}[P]),
\end{align*}
where we use Wald's identity \cite{gallager2012discrete} under condition $\tilde{n}>0$.  
In view of these observations, we evaluate the following:
\begin{align*}
\mathbb{E}&[X_iT_i|K_{i-1}=(Id)] \\ &= \mathbb{E}[(P_{i-1} + T_{o} + I_i)(P_i + S_i)] \\ &\  +\mathbb{E}[(P_{i-1} + T_{o} + I_i)W_i \mathbbm{1}_{T_O + I_i + P_i < S_i}] \\ &= \mathbb{E}[(P_{i-1} + T_{o} + I_i)(P_i + S_i)] \\ &\  +\mathbb{E}[(P_{i-1} + T_{o} + I_i)W_ie^{-\mu (T_O + I_i + P_i)}] \\ &= \mathbb{E}^2[P] + (\frac{1}{\lambda} + \frac{1}{\mu} + T_{o})\mathbb{E}[P] + (\frac{1}{\lambda}+ T_{o})\frac{1}{\mu} \\ &\ + \frac{\lambda (\lambda + \mu)(\mathbb{E}[P]+T_{o}) + \lambda}{(\lambda + \mu)^2} e^{-\mu T_{o}}\mathbb{E}[W_i]M^{P}(\mu).
\end{align*}\vspace{-0.2in}

\subsubsection{$\mathbb{E}[X_iT_i \ | \ K_{i-1}=(B)]$}

In this case, packet $i-1$ finds the second queue in (B) state. $X_i = P_{i-1} + T_{o} + I_{i}$ and if $T_{o} + I_i + P_i > R_{i-1} + S_{i-1}$, $R_{i-1}\leq \tau$ or if $\tau <R_{i-1} \leq T_{o} + I_i + P_i$, then $T_i = P_i + S_i$. On the other hand, if $T_{o}+ I_i + P_i < R_{i-1}$ or $T_{o} + I_i + P_i < R_{i-1} + S_{i-1}$ and $R_{i-1}<\tau$, then $T_i = P_i + W_i + S_i$ where $W_i$ is the wait time in the data buffer. Note that the distribution of $W_i$ is as in in (\ref{frm}) with the same $\mathbb{E}[W_i]$. We next evaluate the conditional expectation:
\begin{align*}\nonumber
\hspace{-0.0in}\mathbb{E}[X_iT_i|K_{i-1}=(B)] & = \mathbb{E}[(P_{i-1} + T_{o} + I_i)(P_i + S_i)] \\ \nonumber  &\hspace{-0.85in} + \mathbb{E}[(P_{i-1} + T_{o} + I_i)W_i (\mu \tau+1) e^{-\mu (T_{o} + I_i + P_i)}], \end{align*} \begin{align*} \nonumber &\hspace{-0.15in} = \mathbb{E}^2[P] + (\frac{1}{\lambda} + \frac{1}{\mu} + T_{o})\mathbb{E}[P] + (\frac{1}{\lambda}+ T_{o})\frac{1}{\mu}  \\  &\hspace{-0.15in} + \frac{\lambda (\lambda + \mu)(\mathbb{E}[P]+T_{o}) + \lambda}{(\lambda + \mu)^2} (\mu \tau + 1)e^{-\mu T_{o}}\mathbb{E}[W_i]M^{P}(\mu).
\end{align*}

\noindent We finally obtain the following:
\begin{align*}\nonumber
\mathbb{E}[X_iT_i]&=\mathbb{E}[X_iT_i|K_{i-1}=(B)]p_{B} + \mathbb{E}[X_iT_i|K_{i-1}=(Id)]p_{I} \\ \nonumber &\hspace{-0.5in}= \mathbb{E}^2[P] + (\frac{1}{\lambda} + \frac{1}{\mu} + T_{o})\mathbb{E}[P] + (\frac{1}{\lambda}+ T_{o})\frac{1}{\mu}  \\ \nonumber &\hspace{-0.5in}+ \frac{\lambda (\lambda + \mu)(\mathbb{E}[P]+T_{o}) + \lambda}{(\lambda + \mu)^2}(p_B\mu \tau + 1)e^{-\mu T_{o}}\mathbb{E}[W_i]M^{P}(\mu),
\end{align*} 
where $p_B$ is as in (\ref{pb}).

\subsection{Average Peak AoI}

We now extend our analysis by following \cite{pimrc19} and obtain $\mathbb{E}[X_{i^*} + T_{i^*}]$ where $i^*$ is the packet index corresponding to the minimum of $X_i + T_i$'s in a given service period. We will use the relation $\mathbb{E}[X_{i^*}+T_{i^*}]=\frac{\mathbb{E}[(X_{i}+T_{i})\mathbbm{1}_{i=i^*}]}{\mbox{Pr}(i=i^*)}$ with $\mathbbm{1}_{i=i^*}$ denotes the indicator function of whether packet $i$ is the minimum index in a service period and $\mbox{Pr}(i=i^*)$ is the corresponding probability. We next consider $K_{i-1}=(Id)$ and $K_{i-1}=(B)$.

\subsubsection{$\mathbb{E}[(X_{i}+T_{i})\mathbbm{1}_{i=i^*} \ | \ K_{i-1}=(Id)]$}

In this case, if $T_{o} + I_i + P_i > S_{i-1}$, then $T_i = P_i + S_i$. If $T_{o} + I_i + P_i < S_{i-1}$, then $T_i = P_i + W_i + S_i$ where $W_i$ is as in (\ref{frm}). Additionally, $\mbox{Pr}(i=i^* \ | \ K_{i-1}=(Id)) = 1$ is easily verified as the next arrival after a packet arriving in (Id) state will be the first among those served together. We then get the following expression:
\begin{align}\nonumber
\mathbb{E}[(X_{i}+&T_{i})\mathbbm{1}_{i=i^*} \ | \ K_{i-1}=(Id)]  \\ &\hspace{-0.5in}=  \mathbb{E}[P_{i-1} + T_o + I_i + P_i + S_i] +\mathbb{E}[W_i \mathbbm{1}_{T_O + I_i + P_i<S_i}], \\ &\hspace{-0.5in}=  \mathbb{E}[P_{i-1} + T_o + I_i + P_i + S_i] +\mathbb{E}[W_ie^{-\mu (I_i + P_i)}],  \\ &\hspace{-0.5in}=2\mathbb{E}[P] + \frac{1}{\lambda} + T_o + \frac{1}{\mu} + \frac{\lambda e^{-\mu T_{o}}}{(\lambda + \mu)} \mathbb{E}[W_i]M^{P}(\mu).
\end{align}

\subsubsection{$\mathbb{E}[(X_{i}+T_{i})\mathbbm{1}_{i=i^*} \ | \ K_{i-1}=(B)]$}

Conditioned on (B) state observed by packet $i-1$, the next packet index $i$ will be the minimum index only if the next packet $i$ arrives after the residual time $R_{i-1}$ (when no packet is sitting at the arrival time of packet $i$) and $R_{i-1}\leq \tau$  (when packet $i-1$ has not been discarded). Recalling that $\tau \leq T_o$, the second condition is sufficient and we have $\mbox{Pr}(i=i^* \ | \ K_{i-1}=(B)) = 1- e^{-\mu \tau}$. In this case (i.e., when $R_{i-1}\leq \tau$), if $T_o+I_i + P_i > R_{i-1} + S_{i-1}$, then $T_i = P_i + S_i$. If $T_o + I_i + P_i < R_{i-1} + S_{i-1}$, then $T_i = P_i + W_i + S_i$ where $W_i$ is as in (\ref{frm}) and we have
\begin{align}\nonumber
\mathbb{E}[(X_{i}+T_{i})&\mathbbm{1}_{i=i^*} \ | \ K_{i-1}=(B)] \\ \nonumber &\hspace{-0.5in}= \mathbb{E}[P_{i-1} + T_o + I_i + P_i + S_i](1-e^{-\mu\tau}) \\ &\ \ +\mathbb{E}[W_i \mu \tau e^{-\mu ( T_o + I_i + P_i)}], \nonumber \\ \nonumber &\hspace{-0.5in}=(2\mathbb{E}[P] + T_o + \frac{1}{\lambda} + \frac{1}{\mu})(1-e^{-\mu\tau}) \\ &+ \mathbb{E}[W_i]\mu\tau \frac{\lambda e^{-\mu T_o}}{\lambda + \mu}M^{P}(\mu).
\end{align}
We finally conclude as follows:
\begin{align*}\nonumber
\mathbb{E}[(X_{i}+T_{i})\mathbbm{1}_{i=i^*}] &= (2\mathbb{E}[P] + T_o + \frac{1}{\lambda} + \frac{1}{\mu})(1 - p_B e^{-\mu\tau})  \\ &\hspace{-0.8in} + \mathbb{E}[W_i](1-p_B + p_B \mu\tau) \frac{\lambda e^{-\mu T_o}}{\lambda + \mu}M^{P}(\mu), \end{align*} \begin{align*}
\mbox{Pr}(i=i^*) &= 1 - p_B e^{-\mu \tau},
\end{align*}
and we combine to get $\mathbb{E}[X_{i^*} + T_{i^*}] = \frac{\mathbb{E}[(X_{i}+T_{i})\mathbbm{1}_{i=i^*}] }{\mbox{Pr}(i=i^*)}$.

\section{Numerical Results}
\label{sec:Numres}

In this section, we provide numerical results for AoI performances focusing on the following four aspects of the system: (i) $T_o$: OFF time duration for the computation server; (ii) $\tau$: Packet deadline to maintain freshness of those taken to the transmission server (recall that $\tau \leq T_o$);  (iii) Processing power $p_c>1$ and the long-term power constraint in (\ref{const}) (iv) $g(.)$: The variation of mean transmission time with respect to computing. We will use $g(\mathbb{E}[P])=B_0 e^{-\alpha \mathbb{E}[P]}$, which represents a convex relation between mean times of computation and transmission with $\alpha$ parameter tuning how fast it affects the outcome. We expect that as $\alpha$ gets larger, it will be more profitable to perform longer computation at the expense of more power spent. We use closed form expressions we derived in the paper while offline simulations verified the expressions. 

We will use Gamma distributed ON-Busy computation time due to its convenience of closed form MGF expression with explicit dependence on first and second moments. In particular, we use $f_P(p)=\frac{k^k \kappa^k}{\Gamma(k)}p^{k-1}e^{-k \kappa p}$ for $p \geq 0$ where $\kappa=\frac{1}{\mathbb{E}[P]}$ with $\mathbb{E}[P]$ representing the mean value and $k >0$ parameter tuning the variance of $P$. This distribution has the following closed form moment generating function (MGF): \[ M^{P}(\mu)=\left(1+\frac{\mu }{k \kappa}\right)^{-k} \]
It is remarkable that we only need the zeroth order MGF to get the average AoI and average peak AoI (in contrast to first and second derivatives needed in closely related work \cite{infocom_arxiv,pimrc19}). 

We view the mean service time $\mathbb{E}[P]$ and the threshold $\tau$ as control knobs to be determined jointly subject to aforementioned system constraints. Typical to this tandem queue model (as in \cite{pimrc19}), we expect to have average AoI and average peak AoI to behave differently and minimizing one may lead to a vastly suboptimal value for the other. To this end, we adopt weighted sum of AoI and average peak AoI as objective:
\begin{align}\label{ths}
\min_{\tau \leq T_o, \mathbb{E}[P] \geq 0, \ \mbox{s.t.} \ (\ref{const})} \omega_1 \mathbb{E}[\Delta] + \omega_2 \mathbb{E}[PAoI]
\end{align}

\begin{figure}[!t]
\centering{
\hspace{-0.6cm} 
\includegraphics[totalheight=0.29\textheight]{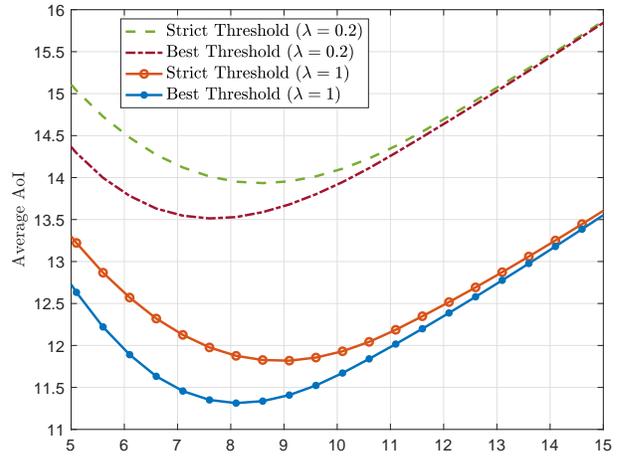}}\vspace{-0.29in}
\caption{\sl Average AoI versus $T_{o}$ for strict threshold and best threshold schemes with $\lambda=0.2$ and $\lambda=1$. }\vspace{-0.29in}
\label{fig:numres1} 
\end{figure}

\begin{figure}[!t]
\centering{
\hspace{-0.6cm} 
\includegraphics[totalheight=0.29\textheight]{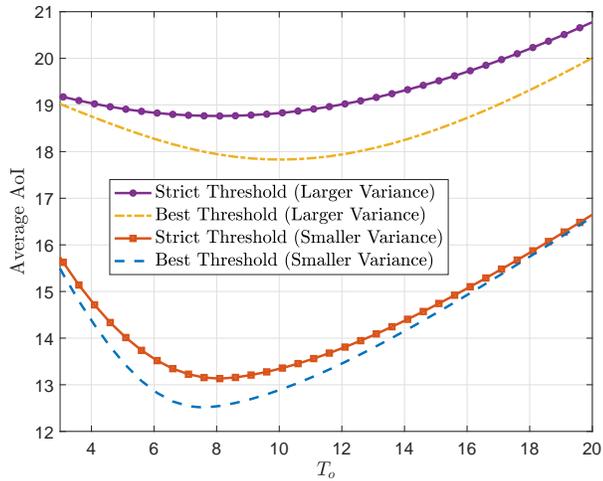}\vspace{-0.15in}}
\caption{\sl Average AoI versus $T_o$ comparing strict threshold and best threshold schemes with smaller and larger variance of compute time $P$. }\vspace{-0.2in}
\label{fig:numres2} 
\end{figure}

We start by setting $B_0=10$, $\alpha=1$, $p_c=10$, $C_{avg}=1$ and $k=0.1$. Fig. \ref{fig:numres1} demonstrates the variation of the average AoI with respect to $T_o$ under $\tau=0$ (the most strict threshold used by the transmission server in which case server drops all packets that arrive in a busy period. This extreme case is equivalent to the transmission server having G/M/1/1 form.) and the best threshold that solves (\ref{ths}) for $\omega_2=0$. It is remarkable that the best selection of $T_o$ that minimizes average AoI is non-zero as keeping the computation server in a stand-by state for some non-zero time duration helps in achieving a better computation-communication tradeoff. We also observe that as the job arrival rate $\lambda$ is increased, the improvement brought by the best threshold becomes more apparent (around $5\%$ for $\lambda=1$). Then, we test the effect of variance of the compute time in ON-Busy on the average AoI by changing parameter $k$ of the Gamma distribution. In particular, we set $k=0.05$ (referred as smaller variance) and $k=0.005$ (referred as larger variance). In Fig. \ref{fig:numres2}, we observe that the larger the variance of computation time in ON-Busy state is, the better the thresholding performs with respect to strict thresholding. However, the value of average AoI increases as a result of increased variance of $P$. This is in line with earlier observations in \cite{pimrc19} and the analysis in \cite{ozel2020} in terms of the variance of computation time. 

\begin{figure}[!t]
\centering{
\hspace{-0.6cm} 
\includegraphics[totalheight=0.29\textheight]{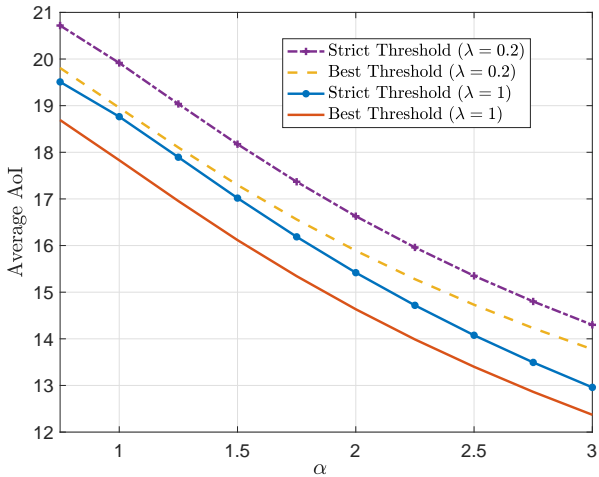}\vspace{-0.15in}}
\caption{\sl Average AoI versus $\alpha$ for strict threshold and best threshold schemes with $\lambda=0.2$ and $\lambda=1$. }\vspace{-0.2in}
\label{fig:numres3} 
\end{figure}

We next consider the effect of $\alpha$ (determining the conversion rate between mean computation time in ON-Busy state and mean transmission time). We use the same parameters except setting $k=0.005$ while $\lambda$ and $\alpha$ are varied. We optimize over $\tau$, $\mathbb{E}[P]$ and $T_o$ numerically for the best threshold scheme whereas $\tau=0$ for strict thresholding. In Fig. \ref{fig:numres3}, we observe that thresholding the wait time of the packet in the transmission side data buffer becomes more beneficial as we increase $\alpha$ (representing the regime of more valuable compute time with respect to transmission time) when optimization is performed jointly over $\tau$ and $T_o$. While average power constraint in (\ref{const}) forces the server to remain idle and hence delegate the status update to the transmitter early, doing so after achieving a reasonable computation and leaving small amount to be transmitted to the receiver (as $\alpha$ is increased) brings noticeable improvement to the AoI performance.

\begin{figure}[!t]
\centering{
\hspace{-0.6cm} 
\includegraphics[totalheight=0.29\textheight]{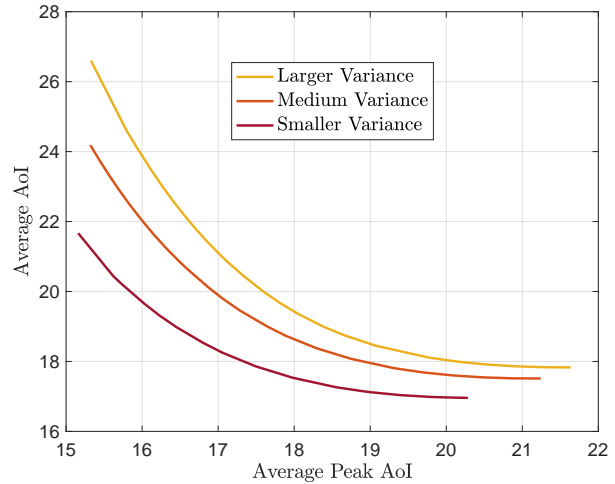}\vspace{-0.15in}}
\caption{\sl Tradeoff curves between average AoI and average peak AoI. }\vspace{-0.2in}
\label{fig:numres4} 
\end{figure}

In other numerical results that we do not report here, we observe similar trends for the variation of average peak AoI with respect to system parameters. To address the conflict between optimizing average AoI and average peak AoI, we run the optimization in (\ref{ths}) for various weights and report the result in Fig. \ref{fig:numres4}. We set $\alpha=1$, $\lambda=1$ and plot for different $k$ values. In particular, we refer to $k=0.008$, $k=0.006$ and $k=0.005$ as smaller, medium and larger variance cases, respectively. We observe that as the variance of computing time is increased, both average AoI and average peak AoI increase. It is noted that the increase in minimum average peak AoI is limited while the sensitivity of the average AoI to a change in variance is notable. \vspace{-0.1in}

\end{document}